\documentclass[aps, prd, twocolumn, lengthcheck, superscriptaddress, showpacs, letterpaper, nofootinbib]{revtex4-1}
\usepackage{amsfonts}
\usepackage{amsmath}
\usepackage{amssymb}
\usepackage{xcolor}
\usepackage{graphicx}
\usepackage{subfigure}
\usepackage{longtable}
\usepackage{slashed}
\usepackage{hyperref}
\usepackage{amsthm}
\usepackage{mathrsfs}
\usepackage{color}
\usepackage{epsfig}
\usepackage{epstopdf}

\def\be{\begin{eqnarray}}\def\ee{\end{eqnarray}}
\def\bi{\bibitem}\def\del{\partial}
\def\la{\langle}\def\ra{\rangle}
\def\be{\begin{eqnarray}}\def\ee{\end{eqnarray}}
\def\lsim{\mathrel{\rlap{\lower3pt\hbox{\hskip1pt$\sim$}}
     \raise1pt\hbox{$<$}}} 
\def\gsim{\mathrel{\rlap{\lower3pt\hbox{\hskip1pt$\sim$}}
     \raise1pt\hbox{$>$}}} 

\allowdisplaybreaks

\begin{document}

\title{Mapping topology to nuclear dilaton-HLS effective field theory\\  for dense baryonic matter}

\author{Yong-Liang Ma}
\email{ylma@ucas.ac.cn}
\affiliation{School of Fundamental Physics and Mathematical Sciences,
Hangzhou Institute for Advanced Study, UCAS, Hangzhou, 310024, China}
\affiliation{International Center for Theoretical Physics Asia-Pacific (ICTP-AP) (Beijing/Hangzhou), UCAS, Beijing 100190, China}

\author{Mannque Rho}
\email{mannque.rho@ipht.fr}
\affiliation{Institut de Physique Th\'eorique, Universit\'e Paris-Saclay, CNRS, CEA, \\  91191 Gif-sur-Yvette c\'edex, France }

\date{\today}

\begin{abstract}
We describe the on-going effort to formulate the baryon-quark continuity in terms of a topology change in the equation of state (EoS) of dense baryonic matter in analogy to -- and inspired by -- the mapping of the characteristics of Chern-Simon topological field theory to Kohn-Sham density functional theory in the fractional quantized Hall effect (FQHE).  This is done by translating the density-dependent characteristics of the skyrmion-half-skyrmion transition formulated in the presence of hidden local symmetry and (hidden) scale symmetry to the density-dependent parameters of a renormalization-group approach to Fermi-liquid fixed point theory. Predictions made in finite nuclei and infinite compact-star matter are presented.
\end{abstract}
\maketitle
\setcounter{footnote}{0}

\section{Introduction}

\subsubsection*{The Problem}

In nuclear processes, broadly two density regimes are  delineated by a possible change of degrees of freedom (DoFs). At low densities up to slightly above the equilibrium nuclear matter density ($n_0\simeq 0.16$ fm$^{-3}$), say, to $n\lsim 2 n_0$, the relevant degrees of freedom are the nucleons $N$ and  pions $\pi$ figuring in what is referred to as ``standard chiral effective field theory" (given the acronym $s$ChEFT) with a cut-off scale set, typically, at $\Lambda_{sChEFT}\sim (400-500)$ MeV. When treated to N$^m$LO for $m=(3-4)$ in systematic chiral power expansion,  {\it ab initio}  calculations in $s$ChEFT have been seen to work highly satisfactorily for nuclear structure in finite nuclei as well as for properties of nuclear matter. This impressive success in nuclear physics can be taken as a proof of Weinberg's  ``Folk Theorem for EFT" as applied to QCD~\cite{FT}\footnote{There are a large number of excellent reviews in the literature, much too numerous, however, to make an adequate referencing. We apologize for not listing them here.}. This $s$ChEFT is expected to be extendable to  $\sim 2$ times $n_0$, but it is most likely to break down beyond it as the relevant degrees of freedom are no longer just the nucleons and pions but more massive degrees of freedom and ultimately the QCD degrees of freedom, quarks and gluons, must enter as density increases. Thus there must be change of DoFs from hadronic to quarkonic at some density. 

Due to the paucity of trustful theoretical tools for guidance in the absence of lattice approach to QCD in dense medium --   in contrast to thermal matter --   there is no clear indication how many and in what form(s)  these changeovers of DoFs  could take place, say,  in the density regime relevant to the center of massive compact stars,  $\sim (5 - 7) n_0$. This presents a totally uncharted domain that could very well encompass several different fields, say, condensed matter, nuclear and particle in addition to astrophysics,

There are several proposals in the literature that invoke quarks in various different forms, perturbative or nonperturbative, to explore this uncharted domain. In this short note we describe a possible strategy that exploits the topological structure of baryonic matter -- without explicit QCD variables -- as density increases beyond $\sim 2n_0$ to access the putative baryon-quark continuity. No explicit quark-gluon DoFs are invoked, but fractional (baryon-charged) quasiparticles induced from topology change are found to play the principal role. The approach is highly unconventional,  certainly unorthodox and much incomplete in various aspects.  Nonetheless it is shown to successfully post-dict not only the properties of nuclear matter in quality more or less comparable to $s$ChEFT at high chiral orders but also accounts for the properties of massive compact stars in fair consistency with what has been established in astrophysical observations.  There is however a surprising new  prediction -- which is extremely simple -- that follows from a topology-change on the sound velocity of the star and its impact on the structure of the core of compact stars. It exposes emergent symmetries hidden at low density.
\subsubsection*{The Motivation}

The basic idea figuring here is largely motivated by the predominant role topology plays in quantum critical phenomena,  beautifully exemplified, among others, by the fractional quantized Hall effects~\cite{tong}.  This idea is germinated by the observation that many-body interactions in strongly correlated condensed matter systems and in nuclear many-body systems, although the basic interactions are different, i.e., QED vs QCD,  intricately  share certain common features. In the former, there have been remarkable breakthroughs by formulating correlated electron problems in terms of topology, thereby mapping many-electron interactions to topological field theories and also the other way around. The strategy we  describe here is prompted to do something analogous to what has been done in the physics of quantum Hall effects (QHE), more or less totally unrecognized in the nuclear community, so there are certain arguments -- and ideas -- borrowed from QHE. Of course they are necessarily of different nature given that we are dealing with strong interactions (QCD) with inherently more complex dynamics than in QED. Furthermore strong interactions have much less access to experiments than in condensed matter systems.
\subsubsection*{The Objective}
Let us first state briefly what the problem is and what the objective of this article is. 

The main stream of current activities in nuclear/astro-physics community motivated by what's heralded as ``first-principle" approaches to nuclear physics and gravity-wave signals of merging neutron stars is anchored on  {\it ab initio} treatments of the $s$ChEFT at high chiral orders and its related density-functional-type theories {\it defined and valid} at the density regime $\lsim 2n_0$. They are then extrapolated, resorting to various sophisticated ``meta-modelings"  relying on the Baysian inference, high-order ``uncertainty  analysis" etc., to the higher densities relevant for massive compact stars for which trustful theoretical tool is at present totally missing.  

The spirit of this article is drastically different from the majority of the approaches pursued in the field. 

The aim of the approach adopted in this note is instead to construct as simple and economical an effective field theory as possible,  implementing what are deemed to be necessary to meet the requirement for the ``Folk Theorem for EFT"  appropriate for normal as well as compact-star matter. The philosophy is then to see how far one can go forward with this extremely simple theory -- never mind the nitty-gritty ``error uncertainties" -- before being hit by ``torpedos."\footnote{In the spirit of Farragut's famous uttering `` Damn the torpedos. Full speed ahead!"} 

In a nutshell, this article attempts to explain why the approach which is perhaps over-simplified and intuitive far from rigor seems to work surprisingly satisfactorily for nuclear dynamics ranging from low density at $n\sim n_0$ where the $s$ChEFT is {\it believed} to be applicable to high density $n\sim (5-7)n_0$ where it is {\it suspected} to break down. 

Although the approach presented here predates the arrival of $s$ChEFT in 1990's, the core idea was the principal theme of the 5 year ``World Class University Project" (WCUP)  established at Hanyang University in Seoul in 2007 funded by  the Korean Government. This WCUP was in some sense in anticipation of the upcoming ambitious Institute of Basic Science (IBS) with the purpose to put Korean basic science on the world's frontier. We will base our discussion on what was initiated in 2007 and continued after the termination of the WCUP/Hanyang up to today.

The results that we will refer to are mostly available in the literature.  The development up to 2018 is summarized in \cite{WS-RM-2018}, written in tribute to Gerry Brown who had made invaluable contributions to the development of theoretical nuclear physics in Korea. More recent developments, which make the story more interesting and more up-to-date, though imbued with a non-negligible dose of speculations, are found in \cite{MR-PPNP} and in arXiv~\cite{MR-manifestation}.  We must state that why this approach works as well as it actually seems to do still remains,  to say the least,  mystifying to us. This most likely accounts for the lack of recognition of the approach by  workers in the field. The objective of this note is to try to demystify it to the best feasible and then speculate in which way the approach can be pushed further to provide justifications.
\subsubsection*{The Strategy}\label{KS}
The best way to motivate the reliance on topology to go from nucleons and pions at low density to quarks (and gluons) at high density is to think of a possible parallel of the approach adopted to how the physics of quantum Hall effects  is formulated in terms of a topological field theory.  What we have in mind in particular is the mapping of the fractional quantum Hall effect (FQHE) given in Chern-Simons topological field theory (CSTFT)  to the Kohn-Sham density functional theory (DFT)~\cite{DF-jain,jain}.  It must be stressed that the parallel cannot be direct, with totally different physics involved, but what figures in  both the FQHE and compact-star matter involves mapping between the {\it microscopic description}, DFT, and the {\it macroscopic description}, CSTFT.  The possible presence of this parallel, although present from the very beginning of the WCU/Hanyang,  was only very recently recognized thanks to the on-going works of string-theory-oriented theorists. It will be suggested that both the KS-FDT-type microscopic approach and the CSTFT-type macroscopic approach figure in the development made at the WCU/Hanyang and since then. 

Let us very briefly describe what the parallel that we see is. 

In \cite{DF-jain,jain}, the system of strongly interacting electrons in the FQHE regime is formulated in terms of composite fermions of electrons bound with even number of quantum vortices involving an $U(1)$ gauge field {\it emergent} from strong correlations of the electrons. The complex effects of many-electron interactions are cast in a single-particle formalism in Kohn-Sham (KS) density functional theory incorporating the emergent $U(1)$ gauge interactions  between {\it weakly interacting}  composite fermions, i.e., quasiparticles,  induced by the quantum mechanical vortices. The gist of the approach then is that {\it the topology of Chern-Simons field theory is translated into the effective field theory,  DFT}.  What's remarkable in this approach is that their DFT does ``faithfully capture the topological characteristics" of the FQHE.

The approach that we will follow is inspired from the analogy of accessing the strongly correlated strong interactions in the density regime $n\gsim (2-4)n_0$ -- where $s$ChEFT is presumably broken down -- to the mapping  of the FQHE to the problem of nearly non-interacting composite electrons in the KS-DFT subject to an emerging ``magnetic field."  This analogy is not totally unfamiliar in its generic form in nuclear dynamics where the effective field theories of QCD, e.g., $s$ChEFT at low density, are extended to higher density regime with the cutoff set higher than $\Lambda_{sChEFT}$ in the form of ``relativistic mean-field theory (RMFT)" with heavier meson DoFs included. In fact one could consider these EFT approaches -- including the G$n$EFT discussed in this note  -- to generically belong to the class of Kohn's DFTs (including KS-DFT) applied in nuclear physics.\footnote{There is in the literature a huge variety of ``mean-field theories," both relativistic and non-relativistic, for treating both finite nuclei and infinite nuclear matter.  Some of them are capable of explaining the ensemble of  available terrestrial and astrophysical observables with success. They remain however mostly phenomenological, having little if any with the fundamental theory QCD, at the densities relevant to massive compact stars.}

We define our principal strategy as follows: Incorporate into an EFT -- that we will call ``G$n$EFT" from here on  to be distinguished from the standard ChEFT --  what are deemed to be ``robust" properties of the dense skyrmion matter built with the DoFs heavier than the pion. This EFT  is to capture as ``faithfully" as feasible the topological characteristics of the skyrmion matter. The basic assumption made is that at high density and in the large $N_c$ limit, the skyrmion matter is a crystal with a {\it negligible} contribution from the kinetic energy term~\cite{crystal}.
 
We mention in anticipation that the $\omega$ meson figuring as the $U(1)$ component of the hidden local symmetry (HLS)  can be identified with the $U(1)$ Chern-Simons field playing a role in the fractional quantum Hall droplet structure for the $N_f=1$ baryon associated with the $\eta^\prime$ singularity. We will return to the $\eta^\prime$ singularity because it is currently argued to be crucial for chiral restoration involving topology at high density, an issue that has only very recently been raised. 

In short, what comes out from the approach treated in this note, extremely simple as it is in concept  -- and highly unorthodox -- in comparison with what's done by others in the field, turns out to work well ranging from nuclear matter density to high density relevant to massive compact stars.  Up to date we see no serious tension with Nature.  
We will limit ourselves to highlight with no details some of the (unorthodx) predictions of the approach.

\section{Topology in baryonic matter}
\subsection{Change of DoFs: Hidden Symmetries}
To simulate the change of DoFs in terms of topology as density goes up from below to above the putative baryon-quark continuity density  denoted as $n_{\rm BQC}$, two hidden symmetries  invisible in QCD in the vacuum ($n=0$) are found to be essential.  One is hidden local symmetry (HLS)~\cite{HLS} and the other is hidden scale symmetry (HSS). The cutoff scale involved for the G$n$EFT should be greater than the cutoff effective for $s$ChEFT. The precise value of the relevant cutoff scale is not needed for what follows, but to be specific, one can take the HLS scale to that given by the vector meson ($V=(\rho, \omega)$) mass $m_V\sim 700$ MeV. The (hidden) scale symmetry is associated with the possible dilaton scalar $f_0(500)$ which will be later considered as a pseudo-Nambu-Goldstone boson of broken scale symmetry.  

As for the baryon-quark crossover density, we will be considering $n_{\rm BQC}\approx (2-4)n_0$ which will be identified later with the topology change density $n_{1/2}$. 

Defining precisely what the hidden symmetries to be incorporated are requires a lot of words, but the basic ideas can be explained rather simply without losing physical content. Here we will summarize the key ingredients that enter in the G$n$EFT.

One of the two symmetries that play a crucial role in the G$n$EFT is the hidden local symmetry (HLS) first formulated in \cite{HLS} and made powerfully applicable to nuclear physics as comprehensively reviewed in \cite{HY:PR}.   An important property of the HLS  concerned is that it is a gauge symmetry {\it dynamically generated} giving rise to a composite gauge field of pions. The existence of such a composite gauge boson is proven to be ``inevitable" if such  a symmetry is implicit in the dynamics~\cite{suzuki} as assumed in our approach. It implies that the vacuum could be tweaked under  extreme conditions, say, by high temperature or high density, such that 
\be
m_V\propto  g_V \to 0\label{vm}
\ee
what is called  the  ``vector manifestation (VM) fixed point"~\cite{HY:PR}.  How to expose such a symmetry at high temperature as in heavy-ion dilepton experiments or at high density as in compact stars is an extremely subtle issue. An important point to note in this connection  is a possible duality \`a la Seiberg between the vector mesons of HLS and the gluons of QCD.

The other hidden symmetry that figures equally importantly is the scale symmetry. There is a long history with a still on-going controversy on how the scale symmetry is manifested in gauge theories, e.g., in strong-interaction physics and in going beyond the Standard Model (BSM). We will eschew going into that issue for which we refer to \cite{MR-PPNP,MR-manifestation} viewed vis-\`a-vis with nuclear physics. As argued there, what is relevant to G$n$EFT is the ``genuine dilaton" scenario (GDS)" of \cite{GDS}.  The GDS posits that there is an infrared (IR) fixed point with $\beta (\alpha_{s\rm IR})=0$ in the chiral limit (with u(p), d(own) and s(trange) quark masses equal to 0) and that the $f_0(500)$ is the scalar pseudo-Nambu-Goldstone (pNG)  boson of spontaneously broken scale symmetry which is also explicitly broken by quantum (scale) anomaly. One of the most distinctive characteristics of this scenario is that the IR fixed point, which is most likely non-perturbative, is realized in the {\it NG mode} with non-zero dilaton condensate (or decay constant) and non-zero pion condensate (or decay constant)\footnote{We note in this footnote that this scenario differs basically from the scenario popular in the BSM circle working with $N_f\sim 8$ in the conformal window of the IR fixed point realizes in the Wigner mode, which is the source of a controversy.}. 
Given that the dilaton $\chi$ is a pNG boson as the pions $\pi$ are and both satisfy soft theorems, one can make a systematic power counting expansion in chiral-scale theory as in chiral theory. The power counting in chiral expansion is well established. Scale symmetry brings an additonal power counting in terms of the expansion of the $\beta$ function. Expanding the $\beta$ function in the QCD coupling $\alpha_s$ near the IR fixed point,
\be
\beta(\alpha_s)=\epsilon \beta^\prime (\alpha_{s\rm IR})+ O(\epsilon^2)
\ee
where $\epsilon=\alpha_s -\alpha_{s\rm IR}$ and $\beta^\prime >0$.  The non-zero mass of $f_0$ is attributed to $ |\beta^\prime (\alpha_{sIR})\epsilon|$. Thus the power counting in the scale symmetry is
\be
\epsilon\sim O(p^2)\sim O(\del^2).
\ee
One can in principle make a systematic chiral-scale counting comparable to chiral symmetry~\cite{GDS,LMR}.   

The matter-free-space mass of the dilaton $\sim 500$ MeV is comparable to that of kaons, so the dilaton is put on the same mass scale as the octet pions. In principle, one has an $SU(3)$ chiral Lagrangian coupled to the dilaton. In the background of nuclear medium, however, as is well known from nuclear phenomena, the lowest-mass meson of the scalar quantum number  is expected to undergo significant  mass-drop whereas kaons do not appreciably. Therefore in medium, identifying the scalar meson to be the dilaton, one can ignore the s quark in doing the chiral-scale dynamics.  Unless otherwise stated, this will be what we will do. We write  the dilaton field as $\chi=f_\chi e^{\sigma/f_\chi}$ which is referred, in the literature, to as ``conformal compensator" field. As written, $\chi$ transforms linearly under scale transformation whereas $\sigma$ transforms nonlinearly. One can use either. In what follows,  we find it more convenient to employ the linearly transforming field $\chi$.  The Lagrangian that combines the pion field $\pi$, the HLS fields $V=(\rho,\omega)$ and the dilaton field $\chi$, suitably written in a chiral-scale invariant way with the scale-symmetry breaking term put in the dilaton potential $V(\chi)$,  will be written as ${\cal L}_{\chi {\rm HLS}}$ with $\chi$ standing for the dilaton.
\subsection{Topology Change}\label{TC}
There is a growing evidence that skyrmions as nucleons could describe finite nuclei and infinite nuclear matter~\cite{multifacet},  but at present it is far from feasible to address dense compact star matter in terms of the pure skyrmion structure. It is however found feasible to extract topological properties of dense baryonic matter by simulating skyrmions on crystal lattice. In doing this, it is assumed that the topological characteristics extracted from the skyrmion crystal can be taken as {\it robust} and could be exploited for making the mapping of topology to density-functional theory analogous to what is described in Sec.~\ref{KS} for the FQHE. Of course considering skyrmions on the crystal  cannot be a good approximation for low-density matter.  For instance it makes little sense to think of nuclear matter, which is best described as a Fermi liquid, as a crystalline. However at high density and in the large $N_c$ limit, baryonic matter could very well be in the form of a crystal~\cite{crystal,dyonicsalt}. It appears quite reasonable -- and it is assumed in this note -- that the baryonic matter at a density greater than $n_{BQC}$ --  the putative baryon-quark transition density -- could encapsulate certain characteristic features of topology that are {\it not} captured in $s$ChEFT-type approaches.

Consider skyrmions constructed with ${\cal L}_{\chi{\rm HLS}}$ put on FCC crystal lattice.  The skyrmions in the system undergo interactions mediated by the DoFs in the way described, e.g.,  in \cite{PV,HLMR}. We list what transpires from the calculation focusing on essentials without going into details .  
\vskip 0.2cm
$\bullet$ {\bf Skyrmion-half-skyrmion ``transition"}
\vskip 0.2cm 

There takes place a topology change from the state of matter with skyrmions to that of half-skyrmions at a density above the nuclear matter density. The transition density $n_{1/2}$ that we equate to $n_{BQC}$ is not predicted by the theory. It turns out from the detailed analysis of the data available by astrophysical observation~\cite{MR-PPNP} that
\be
 2\lsim n_{1/2}/n_0\lsim 4.\label{upper}
  \ee 
In the discussion that follows, we will take this range in mind.   It is plausible that further development in astrophysical observations, e.g., the maximum mass of compact stars, might increase the upper limit of $n_{1/2}$ above what's given in (\ref{upper}).
  
A characteristic feature of this transition is the resemblance to the pseudo-gap phenomenon in superconductivity.\footnote{Possible pseudo-gap phase was also discussed at high temperature~\cite{zarembo}.} The quark condensate $\Sigma\equiv \la\bar{q}q\ra$, identified in the absence of baryonic matter background as the order parameter, goes to zero when space averaged (denoted $\bar{\Sigma}$), whereas the pion decay constant $f_\pi$  remains nonzero. Thus the changeover is not a phase transition in the Landau-Ginzburg-Wilson-type sense. For lack of a better term we will continue to refer to it as ``transition" unless otherwise noted. This feature will play the crucial role in formulating G$n$EFT in the class of field theoretic density-functional approach .
\vskip 0.2cm
$\bullet$ {\bf  Soft-to-hard transition in the equation of state} 
\vskip 0.2 cm

One of the most important observations in the skyrmion-to-half-skyrmion (SHS) transition is the cusp at $n_{1/2}$ in the symmetry energy $E_{sym}$ of the system. The symmetry energy, the coefficient of the term proportional to $\zeta^2=\big((N-P)/(N+P)\big)^2$ in the energy per particle of the system $E(n,\zeta)$,  plays the key role in neutron stars with large excess of neutrons. The $E_{sym}$ decreases as it approaches $n_{1/2}$ from below in density, so providing attraction and then after the cusp at $n_{1/2}$, increases rapidly,  thus giving repulsion. Thus the cusp provides the main --  as it turns out --mechanism for the EoS going from soft-to-hard at $\sim n_{1/2}$. This feature  will be found crucial for the maximum mass of neutron stars as well as certain gravity-wave signals coming from merging neutron stars. What's given in the skyrmion-crystal simulation is, roughly speaking, a mean-field effect and correlation-fluctuations above the mean field would largely smoothen the cusp, but  the soft-to-hard effect remains unaffected in the EoS. It has been shown that this topological feature can be translated into the nuclear tensor force in  G$n$EFT,  reproducing precisely the cusp structure~ \cite{sym-LPR,HLMR,MR-manifestation}.  As an important  consequence of this cusp structure,  is that in going up in density from $n_{1/2}$,  the HLS gauge coupling constant $g_\rho$  is forced to move toward the ``vector manifestation (VM)" fixed point~\cite{HY:PR} at the density $n_{\rm VM}$ at which the vector meson mass vanishes~\cite{suzuki}
\be
m_\rho\sim g_\rho\to 0\ {\rm as}\ n\to n_{\rm VM}\gsim 25 n_0.
\ee
\vskip 0.2cm
$\bullet$ {\bf  Parity-doubling}
\vskip 0.2cm

 The topology change exposes the emergence of another hidden symmetry in strong interactions, namely, the parity doubling. At high density $\gsim n_{1/2}$, the effective nucleon mass deduced from the effective skyrmion mass tends to converge to  a chiral-invariant mass $m_0\sim (0.6-0.9)m_N$
\be
m_N^\ast\to f_\chi^\ast \sim m_0\not\to  0 \ {\rm as}\  \bar{\Sigma}  \to 0.\label{PD}
\ee
It will be seen below that this implies that the trace of the energy-momentum tensor (TEMT)  becomes  -- in the chiral limit -- a function {\it solely} of $f_\chi\sim m_0$ {\it independent of density} at some density $> n_{1/2}$. This will have a crucial impact on the sound velocity of the massive stars. This symmetry is not {\it explicit} in QCD, so one could say it is also emergent.
\vskip 0.2cm
$\bullet$ {\bf  Quasi-free composite fermions}
\vskip 0.2cm

Though not proven yet,  it appears most likely that the parity-doubling structure described above is closely related to  that the ``quasi-fermion"\footnote{As will become clear, the fermionic object concerned later can be classed neither as a pure baryon nor as a pure quark. In the absence of a better name, let us just, for simplicity, call it  ``quasi-fermion."}  in the half-skyrmion phase is a nearly non-interacting quasiparticle. The half-skyrmion is attached with a monopole associated with a hidden $U(1)$ gauge field (say, the $\omega$ field in HLS) whose energy diverges when separated from the other half-skyrmion, but the divergence gets cancelled when the two half-skyrmions are bound, or more precisely, confined~\cite{cho}. The resulting ``composite fermion" made up of two half-skyrmions in the crystal -- with the kinetic energy suppressed  -- seems to resemble the composite fermion in the FQHE with the  kinetic energy absent in the limit of large magnetic field~\cite{jain}. Indeed the skyrmion analysis on the crystal lattice verifies that the quasi-fermions  behave scale-invariantly when the lattice size -- related to the density -- is varied in the half-skyrmion regime $n\gsim n_{1/2}$~\cite{atiyah-manton,PKLMR}.  More on this issue below.

Pushing further the analogy to the FQHE, one may wonder whether the anyonic structure encountered in the FQHE as discussed in \cite{jain} has any relevance in the present problem. Indeed there are observations in skyrmion physics obtained with powerful mathematical techniques that there can be $1/q$ (baryon-)charged  objets with $q$ odd integer~\cite{canfora}, and even other more exotic varieties. We will return to this issue in the second part of this note where fractional quantum Hall droplets or pitas could figure at high density.

\section{Translating topological inputs into effective field theory G$n$EFT}
Given the topological inputs extracted from the skyrmion-half-skyrmion transition, the next step is to incorporate them into the EFT.  To build the EFT concerned as an ``analog" to the KS-DFT in the FQHE in \cite{jain} in the strategy to map topological properties to an EFT, baryons need to be introduced, suitably coupled chiral-scale symmetrically, to ${\cal L}_{\chi{\rm HLS}}$. Let us denote it  ${\cal L}_{{\psi\chi}{\rm HLS}}$ with $\psi$ standing for baryons.  In the presence of explicit baryons, the Wess-Zumino terms in the mesonic Lagrangian ${\cal L}_{{\chi}{\rm HLS}}$ from which the skyrmions are built are absent. This is of course familiar in $s$ChEFT with the standard (nuclear) chiral Lagrangian with or without strangeness where the Wess-Zumino term is absent. 
\subsection{Scale-HLS Baryon Lagrangian in Medium}
Given  one unique Lagrangian defined {\it only} with the {\it relevant hadronic} variables ${\cal L}_{{\psi\chi}{\rm HLS}}$, the topology change is to be encoded in the parameters of the Lagrangian ${\cal L}_{{\psi\chi}{\rm HLS}}$ that change at the density $n_{1/2}$.   Below the transition density, the Lagrangian endowed with the well-defined scaling of the pion and dilaton decay constants as dictated by the matching with QCD -- via correlators as -- proposed a long time ago~\cite{BR91} should reproduce $s$ChEFT.  How well the predicted results fare with the established data of nuclear matter  is widely reviewed (e.g., \cite{MR-PPNP}).    Of course with the extreme simplification, one cannot hope to match the precision enjoyed by $s$ChEFT at high chiral orders.  There  is no surprise here. However the property of $E_{sym}$ approaching $n_{1/2}$ from below does have certain potentially nontrivial features associated with the tidal polarizability (TP) measured in the recent gravity waves which differs from the prediction of $s$ChEFT. We will return to this matter later. 

As stated the parameters of ${\cal L}_{{\psi\chi}{\rm HLS}}$ are drastically affected by the topology change at $n_{1/2}$. They can differ qualitatively in the half-skyrmion phase from what was predicted in \cite{BR91} and hence from naive extrapolation in $s$ChEFT. The most important impact  is on the property of the bound half-skyrmions behaving as a scale-invariant quasi-fermion described in Sec. \ref{TC}.  It is represented in the Lagrangian ${\cal L}_{{\psi\chi}{\rm HLS}}$ as an effective baryonic field with its physical properties -- the mass and coupling constants -- dictated by the topological properties of the half-skyrmion phase, i.e., the near density independence of the effective mass and coupling constants and suppressed kinetic energy.  Equally important is the (assumed) composite (HLS) gauge symmetry \`a la Suzuki theorem with the VM fixed point with the vanishing vector mass at $n_{VM}\gsim 25 n_0$. It makes the vector field coupling to the quasi-fermion strongly weakened, leading to the emergence of (pseudo-)conformal symmetry as explained below.
%
It should be possible to develop chiral-scale symmetric EFT in a parallel to the $s$ChEFT highly successful in nuclear physics at density $\sim n_0$. The cutoff could be set at $\Lambda_{Gn{\rm EFT}}\gsim m_V$.  That this should be feasible in HLS was already discussed in \cite{HY:PR} and initiated in $\chi$HLS~\cite{LMR}. This program of formulating the systematic series expansion is regrettably yet to be formulated. Here we will describe what is feasible in the absence of such a formulation. That is to resort to a strategy resembling what is done in the FQHE~\cite{jain}, a sort of an application to dense nuclear matter of  density functional theory consistent with Hohenberg-Kohn  theorem~\cite{Hohenberg-Kohn}. 

In a nut-shell, the chain of reasonings goes as follows.

As well recognized in nuclear theory circles, the relativistic mean field theory (RMFT) as first formulated in Walecka's linear model~\cite{walecka} belongs to the class of density functional approach. It has been extensively exploited in terms of the Kohn-Sham density functional  in nuclear structure studies. Furthermore in conjunction with $s$ChEFT, {\it ab initio} calculations in Kohn-Sham density functional are being explored, with the possibility of doing precision nuclear structure calculations.  All these efforts are however limited at present to the density regime $\lsim n_{1/2}$.

It is also known, though perhaps not so widely,  that the Walecka model captures Landau Fermi-liquid theory~\cite{matsui}. Next a chiral Lagrangian implemented with the HLS mesons is established~\cite{friman-rho} to lead to the Wisonian renormalization-group approach to Fermi-liquid fixed-point theory~\cite{shankar,polchinski}. It follows then that the Lagrangian  ${\cal L}_{{\psi\chi}{\rm HLS}}$ with the parameters encoding the topology change and matched to QCD in medium  is expected to give in the mean field a highly powerful fermi-liquid fixed point theory that can access densities $\gsim n_{1/2}$. The reliability of the large $\bar{N}$ approximation in the Fermi-liquid fixed point approximation indicates the validity of the mean field for densities near and higher than $n_0$ and more specially for high density~\cite{walecka}. In fact it  is feasible to go beyond the mean-field by including $1/\bar{N}$ corrections\footnote{Where $\bar{N}=k_F/(\Lambda_{FS}-k_F)$ with the cutoff on top of the Fermi surface.} taken into account in what is known as $V_{lowk}$RG. 

The ultimate goal of G$n$EFT is to access the high density regime relevant to massive compact stars. Without going into details here (readily found in the reviews, such as \cite{MR-PPNP}), let us just mention one case at low density $\sim n_0$ which shows a predictive power not shared (up-to-date) by $s$ChEFT which has a close link to what takes place at $n\gsim n_{1/2}$. It is found to provide a simple and potentially elegant resolution to a long-standing mystery lasting several decades of the quenched $g_A$ observed in nuclear Gamow-Teller transitions in light nuclei~\cite{MRgA}.  The key element of the solution is that at the mean field level, i.e., at the Fermi-liquid fixed point, the superallowed Gamow-Teller transition described in G$n$EFT is given by the soft-pion and soft-dilaton theorems involving the density-dependent  Landau fixed-point $F_1$ parameters (that enter in the Landau effective mass). What is particularly interesting is that  the pseudo-conformal symmetry hidden at low density that is responsible for the so-called ``quenching" of $g_A$ seems to emerge in the conformal sound velocity $v_s^2/c^2=1/3$ in massive compact stars as discussed below. 

We should mention as a challenge to the aficionados of what is heralded as ``first-principle" nuclear calculations that this $g_A$ problem could be addressed by the {\it ab initio}  KS-density functional approach. It would be extremely interesting to see whether such a calculation ``torpedos" the G$n$EFT result.   
\section{Going toward massive compact-star matter}
Assuming that the notion of Fermi-liquid fixed point applies to the density regime $n\gsim n_{1/2}$ as it does at $n< n_{1/2}$, one can make two simple calculations of the EoS to go toward the density relevant to massive compact stars. Again given the paucity of the trustful knowledge about the structure of the state involved, the only guidance at hand is the presumed constraints provided by the symmetries assumed to underlie the dynamics. Here they are the hidden local symmetry (of the composite gauge symmetry~\cite{suzuki}) with the vector manifestation (VM) density $n_{\rm VM}\gsim 25 n_0$ and the scale symmetry with the IR fixed point at $n_{\rm sIR}$ which could coincide with $n_{\rm VM}.$\footnote{In the present framework there is no reason to believe that $n_{\rm VM}\simeq n_{\rm sIR}$. The only thing one can say -- and assumed here -- is that both $n_{\rm (VM,sIR)}$ are higher than what's relevant to the maximum density supported by massive compact stars stable against gravitational collapse.} 
\subsection{Dilaton-Limit Fixed Point (DLFP)}
Consider  ${\cal L}_{{\psi\chi}{\rm HLS}}$ for $n\geq n_{1/2}$ with the parameters of the Lagrangian taken arbitrary and completely unconstrained by the topology change, while possibly density-dependent. Assume that the mean-field approximation holds at $n\gsim n_{1/2}$ in the sense defined in the large $N_c$ and large $\bar{N}$ limit. Now what we would like to do is to take what corresponds to going toward the IR fixed point of the GDS.  This can be done by following  Beane and van Kolck~\cite{bira}: First do the field re-parametrization  ${\cal Z}=U\chi f_\pi/f_\chi=s+i\vec{\tau}\cdot\vec{\pi}$ in ${\cal L}_{{\psi\chi}{\rm HLS}}$, have the Lagrangian treated in the mean field  and take the limit ${\rm Tr} ({\cal Z}{\cal Z}^\dagger)\to 0$. This limit is referred to as ``dilaton-limit fixed point" (acronymed DLFP). Two qualitatively different terms appear from this manipulation:  one is regular and the other singular in the limit. The singular part is of the form
\be
{\cal L}_{\rm sing} &=& (1- g_A){\cal A} (1/{\rm Tr} \big({\cal Z}{\cal Z}^\dagger)\big)\nonumber\\ 
&+& (f_\pi^2/f_\chi^2 -1) {\cal B} \big(1/{\rm Tr} ({\cal Z}{\cal Z}^\dagger)\big).
\ee
The first (second) term is with (without) the nucleons involved. The requirement that there be no singularity leads to the dilaton-limit fixed point (DLFP) constraints
\be
g_A\to g_A^{DL}=1\label{gAto1}
\ee 
and 
\be 
f_\pi\to f_\chi \neq 0.\label{fpifchi}
\ee
We have denoted the $g_A$ arrived at in the dilaton-limit fixed point as $g_A^{\rm DL}$ to be distinguished from the $g_A^L$ in \cite{MRgA} arrived at the Landau Fermi-liquid fixed point at $n\sim n_0$. It turns out that the $\rho$ meson decouples from fermions in dense matter even though the gauge coupling $g_\rho\not\to 0$~\cite{PKLMR}.  Therefore the $\rho$ drops out before reaching the VM fixed point.

These ``constraints" are the same as what are in the genuine dilaton properties approaching the IR fixed point~\cite{GDS}. This suggests that  the topological characteristics of the half-skyrmion phase are consistent with the GDS.  We should mention that there are other constraints in dense matter associated with  the DLFP, among which highly relevant to the EoS at high density is the ``emergence" of parity doubling in the nucleon structure discussed above.   
\subsection{Emerging Pseudo-Conformal Symmetry}
Next consider ${\cal L}_{{\psi\chi}{\rm HLS}}$ with its parameters constrained by the skyrmion-half-skyrmion topology change at $n_{1/2}$. As shown, in the half-skyrmion phase,  as the system flows toward the IR fixed point (or  perhaps equivalently the VM fixed point) -- although $n_{1/2}\ll n_{\rm IR} \sim n_{\rm VM}$, the parameters set in {\it precociously} as
\be
f_\pi\to f_\chi,
\ee 
and
\be
\ m_N\to f_\chi\propto \kappa \la\chi\ra \to m_0\label{mN*}
\ee
with $\kappa$ a constant associsted with the parity doubling (\ref{PD}).

In the mean field approximation (which, it should be recalled, corresponds to the Landau Fermi-liquid fixed point), the energy-momentum tensor is easily calculable. It comes out to be~\cite{PKLMR}
\be
\la\theta^\mu_\mu\ra=4V(\la\chi\ra) -\la\chi\ra\frac{\del V(\chi)}{\del\chi}|_{\chi=\la\chi\ra}\neq 0
\ee
where all the conformal anomaly effects including quark mass terms are lumped into the dilaton potential $V(\chi)$. 
Thus one has $\la\theta_\mu^\mu\ra$ as a function of only $f_\chi$ via (\ref{mN*}). It follows that
 \be
 \frac{\del}{\del n} \la\theta_\mu^\mu\ra=0
 \ee
and hence
\be
 \frac{\partial \epsilon(n)}{\partial n}(1-3v_s^2)=0
\ee
where $v_s^2=\frac{\del P(n)}{\del n} (\frac{\del\epsilon}{\del n})^{-1}$ is the sound velocity and $\epsilon$ and $P$ are respectively the energy density and the pressure. It is fair to  assume that there is no Lee-Wick-type anomalous nuclear state at the  density involved, so $\frac{\partial \epsilon(n)}{\partial n}\neq 0$. Therefore we have
\be
v_{pc:s}^2/c^2\approx 1/3.\label{pcs}
\ee 
Note that this is not to be identified with the ``conformal sound velocity" $v_s^2/c^2=1/3$.  The trace of the energy-momentum tensor is not zero, so we call this pseudo-conformal (PC) sound velocity. 

We should mention at this point a surprising observation  in the $V_{lowk}$RG calculation with higher $1/\bar{N}$ corrections taken into account in $E/A$  of the $A$-nucleon ground state. The emergence of the PC symmetry appears to be intricately tied to whether the VM fixed point is lodged in the vicinity of the core of massive stars or at a much higher density. If it is taken far above the core density, say, at $n_{\rm VM}\gsim 25 n_0$, then $E/A$ at $n\gsim n_{1/2}$  can be very accurately reproduced by the two-parameter formula~\cite{PKLMR,MR-PPNP}
\be
E/A= -m_N + X^\zeta (n/n_0)^{1/3}+Y^\zeta (n/n_0)^{-1}\label{E/A}
\ee 
where $\zeta=(N-Z)/(N+Z)$ and $X$ and $Y$ are the constants to be fixed by equating (\ref{E/A}) to the $E/A$ given in $V_{lowk}$RG at $n=n_{1/2}$ by continuity in the chemical potential and pressure.  $X$ and $Y$ depend on where $n_{1/2}$ is located. One can show that $\la\theta_\mu^\mu\ra$ is a constant independent of density for any values of $X$ and $Y$. This then gives rise to the PC sound velocity (\ref{pcs}). On the contrary if if $n_{\rm VM}$ were taken at $n\sim 6n_0$, say, in the center of massive stars, then the sound speed $v_s^2/c^2$ could not set in at the PC value in the range of star density but would exceed 1/3. {\it This strongly suggests that the PC symmetry may be intimately tied to the VM property of hidden local symmetry.}

Note that (\ref{pcs}) is an approximate equality (with non-vanishing TEMT), not an equality which would hold at the asymptotic density $\gg n_{1/2}$.  In the density regime concerned, $n\lsim 7 n_0$, there can however be deviations due to the quark mass term,  and also higher-order terms of the anomaly-induced symmetry breaking involving the anomalous dimension $\beta^\prime$ that would make the sound speed deviate from (\ref{pcs}).

What is most glaringly different between the prediction of the G$n$EFT and that of {\it all} other models in the literature is the onset of the pseudo-conformal (PC) sound speed (\ref{pcs}) at a relatively low density $\sim 3n_0$ and stays more or less constant up to the central density $\sim 6n_0$ of massive star $M_{max}\lsim 2.3 M_\odot$.  

As far  we know, there are no observables so far measured with which the results of the G$n$EFT (including the recent GW observables)~\cite{MR-PPNP} are odds. Because of the change of parameters of the Lagrangian  ${\cal L}_{{\psi\chi}{\rm HLS}}$ controlling the EoS obtained from G$n$EFT, the most drastic of which is the cusp in $E_{sym}$ at the leading order, there occur strong fluctuations in the density regime $\sim (2-4)n_0$ at which the topology change takes place. This gives rise to a spike in the sound velocity in that region after which the sound velocity $v_{pcs}^2/c^2$ stabilizes  quickly above $\sim 3n_0$. The strength of the spike below  the transition region can vary depending on the value of $n_{1/2}$. It can even overshoot the causality limit, for instance,  for $n_{1/2} \gsim 4 n_0$. This strong enhancement in the sound speed going over the normal hadronic-to-non-hadronic crossing can also be seen with the transition mediated by hadronic-quarkyonic continuity~\cite{quarkyonic}. Thus this aspect of the sound speed could very well depend on how the change-over from hadronic to other forms of the state of matter takes place.  This of course would be a too difficult an issue to accurately sort out in the (over)simplified description. What is less unambiguous is the precocious onset of the PC sound velocity.  

The robust takeaway from this result is that in the way the PC symmetry permeates from low density in the $g_A^L\approx 1$ to high density in the $g_A^{\rm DL}=1$,  the PC sound velocity simply reflects the precocious emergence of the same PC symmetry. Among others it predicts that in the core of massive stars at a density $\sim 6 n_0$,  the objects found there are the composite quasi-fermions of bound half-skyrmions.  

The question then is: Are these quasi-fermions unrelated to what might be described as ``deconfined quarks"?  

It has recently been argued in \cite{deconfined}, based on detailed analyses combining astrophysical observations and  theoretical calculations, that the matter in the core of maximally massive stars exhibits the characteristics of  ``deconfined phase" and suggests that the fermions residing in the core are most likely ``deconfined quarks."  The prediction of G$n$EFT differs from this interpretation.  The result of this note is that the objects found in the core are neither purely quarks nor purely baryons but quasi-fermions of the confined half-skyrmions~\cite{core}. The resemblance is however uncanny if one compares the predictions $P/\epsilon$ where $P$ is the pressure and $\epsilon$  is the energy-density as function of density $n$ and the polytropic index $\gamma=d({\rm ln}P)/d({\rm ln}\epsilon)$ made in the description given above with the analysis of \cite{deconfined}. This, we suggest, is the reflection of the topology change, a.k.a., baryon-quark continuity.  We now turn to this issue in terms of what might be called ``hadron-quark duality."
\section{Hadron-Quark Duality\\ as a Cheshire Cat Phenomenon}
The description given above involves composite fermions made of half-skyrmions in some sense ``masquerading" as fractionally charged quarks.    As noted above, however, half skyrmions are not the only objects that skyrmions can turn into. There could be other fractional objects such as mentioned  in \cite{canfora} and others to be mentioned below.

The crucial question in confronting the theory with experiments such as \cite{deconfined} is how what appear to be ``deconfined quarks" are captured by what appear to manifest as pseudo-quarkonic suggested in this note.  Consider $N_c=3$ ``confined quarks"  in, say, the MIT bag for a nucleon  in nuclear matter of mass number $A$ . When squeezed in dense matter, as the bags overlap, one can visualize the quarks ultimately percolating from one bag to another bag and then coalesce into one big bag of $N_c\times A$ quarks. In 1970s, this is the way some nuclear theorists thought of the $^{208}$Pb nucleus as 624 quarks interacting via perturbative QCD  confined within a giant bag. Such a picture was not -- and still is not -- a feasible one for the reason by now well-known at least for low density. Even so, incorporating the MIT bag structure with asymptotic free interactions at high density could make at least qualitative sense at asymptotic densities.  Indeed many papers have been written where low-density hadronic description is hybridized with MIT bag description at increasing density.  They typically involve phase transitions. Let us cite just a couple of the most recent of them here \cite{MIT} where other relevant references can be found.

One possible alternative was put forward for nuclear dynamics at low energy (and low density) by what was called by ``Cheshire Cat Principle"~\cite{CCP} whereby $N_c$ quarks in a bag transform into a topological soliton, skyrmion, so the quarks disappear into the ``smiles" of the Cheshire Cat with the solitons interacting via fluctuating meson exchanges, in the way Weinberg admitted  as what ``nuclear physicists knew what they were doing" before the advent of the $s$ChiEFT as prescribed in the Folk Theorem.
\subsection{``Infinite Hotel" for $N_f\geq 2$: Skyrmions}
What takes place can be imagined as a quark in a ``jail" trying to escape from the jail, fully occupied,  like the filled Dirac sea\footnote{This ``jail-break" scenario is beautifully described in \cite{jail-break}. Actually $N_c$ quarks are involved but we focus on only one of them.}.  A massless quark swimming on top of the sea, say,  to the right  in one spatial dimension\footnote{The argument can  be straightforwardly extended to 3 spatial dimensions.},  in an attempt to escape the jail,  gets blocked at the ``jail wall," so is unable  to escape.  It cannot swim back on top of the Dirac sea,  because chiral symmetry forbids it. But it can plunge into the Dirac sea which is feasible, because the Dirac sea is infinite, and swim back to the left inside the sea. This infinite Dirac sea can be likened to an  ``infinite hotel (IH)"~\cite{jail-break}.  This exploitation of the infinity  is  a quantum effect known as ``anomaly" in gauge theory.

There is one serious problem in this scenario, however. The fermion (baryon) charge carried by the quark disappears into the Dirac sea, so  the baryon number is apparently ``violated" in the process.  In QCD, the baryon charge is absolutely conserved, so the fermion charge cannot disappear. Here takes place a miracle.  The fermion charge is relayed to the ``pion"  that clouds the outside wall,  with the pion (boson) turning into a baryon (fermion). This is by now the well-known story of skyrmions in (3+1)D mathematically characterized by the homotopy group $\pi_3 (S^3)=\mathcal{\cal Z}$ for the $N_f\geq 2$ systems.

This IH phenomenon can be considered to involve two domains, one the quark-gluon one and the other the hadronic one. There are two modes of a global symmetry, i.e., chiral symmetry, involved:  Wigner-Weyl (WW) mode inside the bag and Nambu-Goldstone (NG) mode outside the bag. Therefore the jail wall can be taken as a thin ``domain wall (DW)" that delineates two vacua. This is the ``jail-break" scenario for the $N_f=2$ (i.e., proton and neutron) case. 

The upshot is that the leaking baryon charge is taken up by the pion as a soliton. So in nuclear physics, we argue that for the given soliton chiral angle  $\theta(R)$, the leaking baryon charge $1-\theta (R)/\pi$ (in 1 spatial dimension) is lodged in the skyrmion cloud while the rest of the  charge  $\theta (R)/\pi$ remains in the bag, yielding the total baryon number 1 for a single baryon.  When  the bag is infinite the whole baryon charge is lodged  inside the bag, while when the bag shrinks to zero size the whole baryon charge goes into the skyrmion cloud. So the size of the bag has no meaning for the property of the quark. The confinement size $R$ is therefore an unphysical quantity. One can think of this process as the pion fields giving rise to the baryons as solitons. This is what is referred to as ``Cheshire Cat Phenomenon" or ``Cheshire Cat Principle (CCP)"~\cite{CCP}. This   is akin to the disappearance of the Cheshire Cat in ``Alice in the Wonderland" with the baryon number playing the role of the cat's smile.  In fact it could be more appropriate to identify this phenomenon as a gauge artifact and formulate a gauge theory for the phenomenon~\cite{CC-gauge}.

This discussion of the CC ``smile" applies straightforwardly to (3+1) dimensions. It has indeed been verified by Goldstone and Jaffe~\cite{GJ}  in terms of the spectral asymmetry $\eta(s)$ (defined in (\ref{SA}) below) which gives the baryon charge lodged inside the bag for a given chiral angle $\theta (R)$. The fractionalization of the baryon charge is exact thanks to the topology involved. In (1+1) D, an exact bosonization allows an in-principle CCP also for non-topological processes.    But in the absence of bosonization, such exact CCP does not exist in nuclear processes  in (3+1)D, so much of what one can say of  the processes in nature that are not topological is at best approximate. 
\subsection{No Infinite Hotel for $N_f=1$ Baryons}
The IH scenario discussed above famously does not work when the number of flavors is one. This is because $\pi_3 (U(1))=0$. One then wonders whether there is no soliton for baryon coming from the flavor singlet meson $\eta^\prime$. This puzzle was recently resolved by ideas developed in  condensed matter physics. It has been suggested by Komargodski~\cite{zohar} that the $\eta^\prime$ can turn into a flavor singlet baryon -- denoted from here on as $B^{(0)}$ -- as a fractional quantum Hall (FQH) droplet. At first sight this  FQH droplet (pancake or pita) is unrelated to the usual skyrmion corresponding to the nucleon. 

There are two questions raised regarding this ``dichotomy"~\cite{dichotomy} between the skyrmions and the FQH droplets. The first is: Is there any relation between the two topological objects, the FQH droplet for $B^{(0)}$ and the skyrmion for nucleons?  The second is: Is the phenomenon of the FQH droplets relevant to the EoS at high density? Both questions are in some sense related.
\subsubsection{Baryon for $N_f=1$}
Let us first discuss whether Komargodski's FQH pancake model can be given a formulation in terms of a Cheshire Cat phenomenon. This rephrases what was done in \cite{MNRZ}. 

Suppose the quark in the bag is of $N_f=1$ in the jailbreak scenario.  Let the quark be coupled at the wall $x=R$ to the flavor-singlet meson $\eta^\prime$. Again the confinement leads to the breaking of the baryon charge and gives rise to an anomaly, but since $\pi_3 (U(1))=0$, it {\it cannot} go into the infinite hotel because the topology does {\it not allow} it.  We will come later to how and why the topology might ``dictate" the flow. So where does it go?
The answer~\cite{MNRZ} is that the quark moving in the $x$ direction is allowed to escape by flowing in the $y$ direction and go into a  2d quantum Hall-type pancake, taking care of the anomaly generated by the boundary condition\footnote{Below this boundary will be identified with a thin domain wall (DW).} and keeping the baryon charge conserved. This is known as the ``anomaly in-flow" mechanism leading to the Chern-Simons topological term ~\cite{callan-harvey},  which in 3-form reads 
\be
\frac{N_c}{4\pi}\int_{2+1} ada\label{cs}
\ee
where  $a_\mu$  is the Chern-Simons field which is to capture strong correlations in QCD -- and will be identified later with the $U(1)$ field in HLS, namely the $\omega$ meson~\cite{karasik1,karasik2}. The Cheshire Cat Principle, if held, would imply  that the baryon charge leaks {\it completely} into the FQH droplet, with the ``smile" reducing  to a $U(1)$ vortex line on the pancake. That the resultant FQH droplet correctly carries the baryon charge  is assured by the gauge invariance of the Chern-Simons term (\ref{cs}). How this comes about can be explained in terms of a chiral bosonic edge mode~\cite{karasik1}.\footnote{This edge mode will be found to play a key role in accessing the EoS for massive compact stars~\cite{MR-PPNP}}  
In accordance with the global symmetries of QCD,   the $B=1$ baryon with $N_c=3$ quarks must then have spin $J=N_c/2=3/2$. This yields the high-spin baryon. Thus when the bag is shrunk to zero size, the Cheshire Cat smile resides in the vortex line in the FQH droplet. For instance for $N_c=3$, this picture yields the $\Delta (3/2,3/2)$.  The same  $\Delta (3/2,3/2)$ also appears in the rotational quantization of the skyrmion with $N_f=2$ which comes from the $\infty$-hotel mechanism which does not work for the $N_f=1$ baryons. These two descriptions present an aspect of the dichotomy problem: Whether or how they are related?
\subsubsection{Baryons for  $N_f\geq 2$}\label{dp}
Instead of a flavor-singlet quark, now consider the jail-breaking scenario of the doublet  u  snd  d quarks. There seems to be nothing to forbid the quark from flowing, instead of dropping into the infinite hotel giving rise to a skyrmion, into the $y$ direction as the flavor-singlet quark did to compensate the anomaly generated by the bag wall. Or is there? This is the question raised. But let us blindly apply the same anomaly-flow argument in  CCP  to the $N_f$-flavored quark. The spin-flavor symmetry for the flavor $N_f\neq 1$ will of course be different. Given $N_f=2 $,  we expect to have a non-abelian Chern-Simons field $\mathbf A_\mu$ in place of the abelian $a_\mu$~\cite{MNRZ},
\be
\frac {N_c}{4\pi}\int_{2+1} {\rm Tr}\left(\mathbb A d\mathbb A+\frac 23 \mathbb A^3\right).
\label{nonabliean}
\ee
This presents an alternative jail-break scenario to the infinite-hotel one. 

But there arises the question: What makes nuclear matter (at $n\sim n_0$) realized as a state of skyrmions as Nature seems to indicate, instead of stacks of fractional quantum Hall pancakes or pitas~\cite{karasik1} or  combinations of the two? Is the non-abelian Chern-Simons droplet a meta-stable state absent at low density but could figure at high density? This question is addressed below somewhat speculatively following the recent developments on the role of two hidden symmetries,  flavor local and scale,  intervening at high density involving fractional quantum Hall droplets~\cite{karasik1,karasik2,kitano}.

\subsection{Fermion Number and Hall Conductivity\\ on Domain Wall}\label{domainwall}
The Cheshire Cat Principle~\cite{CCP} posits that physics should not depend on confinement size (such as the MIT bag radius $R$). It could be phrased even as a  gauge dependence in gauge theories~\cite{CC-gauge}. This CCP on the confinement size  was proven by showing that the baryon charge does not depend on  $R$~\cite{CCP}. This follows from that  the baryon charge is topological.  In (1+1)D, the existence of the exact bosonization allows  CCP for also non-topological quantities. In (3+1)D, there is no exact bosonization and hence there is no exact CCP for other than the baryon charge although approximate CCP holds for certain quantities like the flavor singlet axial charge of the proton~\cite{FSAC}.  
In \cite{MNRZ}, the CCP was established also for the $N_f=1$ baryon for the baryon charge with the fractional quantum Hall droplet replacing the skyrmion  for $N_f\geq 2$.\footnote{In \cite{MNRZ}, the bag boundary is taken as a domain wall (DW). Whether the bag boundary can indeed be thought in terms of a DW is not clear and remains to be scrutinized in detail. In modern developments in gauge theories,  the concept of DW (together with ``interface") plays a singularly important role. This is particularly so, in particular in QCD, and is a huge subject in the literature.  We insert a (much too) brief comment as a footnote just to give an idea. The $\theta$ dependence in QCD with massless quarks makes the pertinent case in this note as will be elaborated below.

As well known, the CP symmetry is  spontaneously broken for the vacuum angle $\theta=\pi$. Suppose  $\theta$ varies from 0 to $2\pi$.  There results a domain wall with Chern-Simons theory on it. Now when quarks are massless, since the bulk property of the theory depends on $m^{N_f} e^{i\theta}$,  the $\theta$ dependence is eliminated, replaced by a shift of $\eta^\prime$.  This is the anomaly cancelation restoring  CP symmetry in 4D. Thus the emergence of $\eta^\prime$ in the problem.}  

In oder to understand what's happening, let us re-derive the CCP result for $N_f=1$ baryon of \cite{MNRZ}. This we will do in (3+1)D by considering the bag boundary as an extremely thin ``domain wall" (DW) located at $x_3=0$. Following \cite{vassi},  we will consider quantized Dirac fermions -- say, ``quarks" -- in interaction with a background $U(1)$ gauge field $a_\mu$, and scalar $\sigma$ and pseudo-scalar $\pi$ fields
\be
\cal{L}=\bar{\psi}{\cal  D}\psi\label{vassi-L}
\ee
with 
\be
{\cal D}=i\gamma^\mu (\del_\mu-ie a_\mu)-(\sigma+i\gamma_5 \pi), \ \sigma^2 +\pi^2=1.
\ee
$a_\mu$,  the $U(1)$ component of HLS,   will be more precisely specified below.

Consider the background fields change rapidly near $x_3=0$ and go to asymptotic values. 
One is interested in the {\it vacuum} baryon number $B$ given by 
\be 
B=-\frac 12\eta(0,H)
\ee
where  $\eta (s, H)$ is the  spectral asymmetry that was computed in \cite{GJ} (for the infinite --hotel scenario)
\be
\eta (s, H)=\sum_{\lambda >0} \lambda^{-s} -\sum_{\lambda<0} (-\lambda)^{-s}\label{SA}
\ee
where $\lambda$ is the eigenvalues of the Dirac Hamiltonian $H$.  
With some reasonable approximations, it was obtained in \cite{vassi} that
\be
B=-\frac{g}{4\pi^2} \theta|^{x^3=+\infty}_{x^3=-\infty}\int d^2x f_{12}
\ee
where $\theta \equiv ({\rm arctan}(\pi/\sigma))$ and $f_{\mu\nu}$ is the gauge field tensor. Note that the vacuum fermion number $B$ has two components, first the Goldstone-Wilczek fractionalized fermion number~\cite{goldstone-wilczek} and the other the magnetic flux through the $(x,y)$ 2-d plane  

Consider next a domain wall background defined by the fields $\sigma$ and $\pi$  that depend on $x^3$ only. The one-loop effective action in the non-static background, in (3+1)D, is found to give the parity-odd action~\cite{vassi}\footnote{Why the parity-odd action becomes relevant is explained below.}
\be
S=\epsilon^{\mu\nu\rho 3} \int d^4x d^4y G(x,y)a_\mu (x)\del_\nu^y a_\rho (y) 
\ee
where $G$ is a complicated non-local function of $x^3$, $y^3$ and $z^\alpha=x^\alpha-y^\alpha$, $\alpha=0,1,2$. In the long-wavelength limit in the form factor $G$,
the action can be written as a Chern-Simons term 
\be
S=g^2 \frac{k}{4\pi} \epsilon^{\mu\nu\rho 3} \int d^3y^\alpha a_\mu(y^\alpha,0)\del_\nu a_\rho (y^\alpha,0)\label{S}
\ee
with
\be
g^2\frac{k}{4\pi} = \int d^3x^\alpha dy^3 dx^3 G(z^\alpha, x^3,y^3).\label{k}
\ee
Here $k$ can be identified as the ``level" in the level-rank duality of the Chern-Simons term. 

At this point one can make contact with what was done in the CCP structure~\cite{MNRZ}. For this consider the domain wall located at $x_3=0$ with the ``quark" modes inside the bag $x_3 < 0$ corresponding to  the Cheshire Cat smile.  The $U(1)$ field in (\ref{vassi-L}) could be considered,  as suggested in \cite{karasik1,kitano},  to be the $\omega$ field when the vector mesons $\rho$ and $\omega$ in HLS are treated as the color-flavor locked $U(N_f)$ gauge fields dual to the gluon fields in QCD~\cite{KKYY}. Then the  $\omega$ field can be taken as the Chern-Simons field that captures \`a la CCP the strongly-correlated excitations outside the bag. Now for $U(N_f)_{-N_c}$ dual to $SU(N_c)_{N_f}$ spontaneously broken, the vortex configurations in three dimensions made up of $\rho$ and $\omega$ carry magnetic and electric charges of $U(1)^{N_f}$. The electric charge in the CS term can then be identified with  the baryon charge~\cite{kitano,KKYY}. This allows one to obtain  the vector current from the action $S$ (\ref{S}), the time component of which is
\be
J^0 (x)=\frac{1}{g}\frac{\delta}{\delta a_0(x)} S.
\ee
The baryon number is~\cite{vassi} 
\be
B=\int d^3x J_0 (x)=\frac{gk}{2\pi}\int f_{12} d^2x.
\ee
Setting the Dirac quantization for the magnetic flux threading the vortex~\cite{KKYY}
\be
\frac{g}{2\pi}\int  f_{12} d^2x =1
\ee
one finds the baryon charge equal to the level
\be 
B=k.
\ee 
This is the baryon charge lodged {\it in the vacuum.}

Now to make the connection \`a la \cite{vassi} to the Cheshire Cat scenario discussed in \cite{MNRZ}, we identify the chiral angle for  $\theta$ which is $ =\eta^\prime/f_{\eta^\prime}$ in \cite{MNRZ}, and impose at $x^3=0$ the Cheshire Cat boundary condition 
\be
(1-i\gamma^3 e^{i\gamma_5\theta})\psi|_{x^3=0}=0. \label{bc}
\ee
Then the change in baryon charge is given by
\be
\Delta B\approx \frac{\Delta \theta}{2\pi} 
\ee
where  $\Delta\theta$ is the jump of the $\eta^\prime$ field across the chiral bag boundary. This is the same result obtained in \cite{MNRZ}. The Cheshire Cat dictates the baryon charge $B_{out}=1-B_{in}$ to be lodged in the Chern-Simons action
\be
S^\prime =g^2 \frac{k^\prime}{4\pi} \epsilon^{\mu\nu\rho 3} \int d^3y^\alpha a_\mu(y^\alpha,0)\del_\nu a_\rho (y^\alpha,0)\label{S'}
\ee
so it must be that
\be
k^\prime=1-k.
\ee

Here are two crucial points, among others, to note. First of all, as pointed out in \cite{vassi},  the Chern-Simons term (\ref{S}) or (\ref{S'}) by itself is  not topological. This is because the level $k$ or $k^\prime$ separately as defined is not an integer so the action is not gauge invariant for $R\neq 0$ or $\infty$. The sum of the baryon charges of inside and outside is required by the anomaly cancellation. This must be related to the color anomaly  found in \cite{colorleakage}. 

Second one could have {\it naively} done the same analysis for the $N_f=2$ case with the pion fields included. That would have given rise to nonabelian CS theory with the same results as in the CCP strategy.  So one is back to the dichotomy problem.

\section{ The Dichotomy Problem}
%

\subsection{Indispensable Role of Vector Mesons}
It is worth recalling that the key ingredient in arriving at the density-functional formalism G$n$EFT that implements the baryon-quark continuity for high density was the hidden local symmetry and the scale/conformal symmetry. What resolves the dichotomy problem is most likely also the working of these two same symmetries emerging from strong correlations  ``dual" to QCD. 

It has recently been realized that if one incorporates the $\eta^\prime$ field in the two-flavor chiral field as
\be
U=\xi^2=e^{\eta^\prime/f_\eta}e^{i\tau_a\pi_a}
\ee
then the ``hidden" Wess-Zumino term in the HLS Lagrangian\footnote{This term corresponds to the ``homogeneous" Wess-Zumino term of \cite{HY:PR} with its (arbitrary) four coefficients  constrained by, e.g., the vector dominance (VD)} unifies the baryon currents for both the FQH droplet for the $N_f=1$ baryons and the skyrmions for the $N_f=2$ baryons~\cite{karasik1,karasik2}\cite{dichotomy}.  As explained by Karasik,  in the effective theory that contains both $\eta^\prime$ and the HLS fields, the $\eta^\prime$ cusp that accounts for the jump from one vacuum to the other at $\eta^\prime=\pi$ does not appear. Thus the emergent theory on the $\eta^\prime$ DW is encoded in the effective theory that contains the HLS fields in the presence of $\eta^\prime$. 

This leads to two observations. 

The first is that the CCP calculation of the nonabelian Chern-Simons action in \cite{MNRZ} has a hidden assumption for the presence of  the $\eta^\prime$ cusp, so the quarks are prevented from falling into the infinite hotel. The same is true with the DW calculation of \cite{vassi} in Sec. \ref{domainwall}.

Secondly one should be able to dial a background parameter (say, density in our case) to go from the skyrmion matter  to the matter that bares the FQH droplet buried in the matter. For this process the dilaton of hidden scale symmetry, not considered in \cite{karasik2},  could be crucial as discussed in \cite{dichotomy}. This is because the emergent field $\omega$ (a.k.a. Chern-Simons field) in HLS is affected by the long wave-length scalar nuclear correlations developed at high density in such a way that the dilaton limit $\bar{\Sigma}\to 0$ suppresses the $\omega$ mass, thereby  making visible  the presence of FQH droplets in the medium. This is a heuristic reasoning as it stands but could be sharpened.

In \cite{kitano}, a scenario different from that of \cite{karasik1,karasik2} is suggested for the role of hidden local symmetry. There the coupling of the Chern-Simons fields in the bulk couple with the edge modes of vector mesons making the vector mesons gauge bosons. At the moment which scenario is preferred is not clear. But what's absolutely clear is that hidden symmetries ``dual" to QCD symmetries (e.g., HLS vector mesons as Seiberg-dual to the gluons~\cite{KKYY}) must be essential for the phase structure at high density, perhaps already at the density relevant to compact stars.
\subsection{Dense Matter as ``Sheets" of Pancakes/Pitas}
%

As noted, at low density, the $N_f=2$ quarks in the bag must be tending to fall into the infinite hotel, hence giving rise to skyrmions in (3+1)D.  This may be driven by the parameters of the Lagrangian that unifies the $N_f\geq 2$ and $N_f=1$ baryons  to have the $B^{(0)}$ effect  {\it suppressed} at low density.  However as density increases,   the parameter change in the scale-symmetric Lagrangian ${\cal L}_{\psi\chi {\rm HLS}}$ that distorts the baryon current from the unified current to the ${N_f=1}$ current could transform the EoS state toward the Chern-Simons FT structure. One possible scenario for this is indicated in the recent skyrmion crystal analyses of dense matter where an inhomogeneous structure is found to be energetically favored over the homogeneous one at high density.  It has been found that dense matter consists of a layer of sheets of ``lasagne" configuration with each sheet supporting half-skyrmions~\cite{PPV}.
%
The constituents of this layer structure are quasi-fermions consisting of fractionalized quasiparticles  of 1/2 baryon charge, appearing in baryon-quark continuity at a density $\sim n_{1/2}$, drastically different from those of the pasta structure discussed for the dilute outer layer of compact stars. In the Skyrme model (with pion field only) used  in \cite{PPV}  the quartic (Skyrme) term effectively encodes massive degrees of freedom,  including the hidden local fields, the monopole structure hidden in half-skyrmions etc. described above.  It appears feasible to formulate this ``sheet dynamics"  by a stack of FQH pancakes or pitas with tunneling half-skyrmions between the stacks, somewhat like arriving at the Chern-Simons structure of FQHE in (2+1)D with a stack of (1+1)D quantum wires~\cite{quantumwires}\footnote{There are many papers on this matter in the literature. A good article with many relevant references is  \cite{quantumwires}}. In the tunneling process it may be possible that the half-skyrmions transform to 1/3-charged quasiparticles resembling  quasi-quarks as discussed in \cite{ vento-half-skyrmion}. On the DW, the 1/3-charged objects could behave as ``deconfined" quarks as discussed in \cite{deconfinedDW}. This could then explain why the composite quasi-fermions given in G$n$EFT behave similarly to the ``deconfined quarks" in the core of massive compact stars~\cite{deconfined}. 
\section{Conclusion}
The question addressed in this note was: Is it feasible with a single unique effective Lagrangian to address the equation of state from normal nuclear matter to massive compact-star matter resorting to only one set of degrees of freedom with the vacuum sliding with density but without  phase transitions ? Put in another way, how far can one go with such a ``unified" formalism without getting into fatal conflict with either empirical or theoretical constraints? 

Influenced by strikingly successful developments in strongly correlated condensed matter physics  together with impact on particle physics, an extremely simplified approach to the EoS for massive compact stars is formulated in terms of topology change to account for possible ``continuity" from seemingly hadronic variables to QCD variables at high density $n_{1/2}\gsim 3n_0$. The single Lagrangian adopted in this study consists entirely of hadronic variables, the pion and the nucleon that figure in the standard nuclear EFT plus the massive degrees of freedom $\rho$, $\omega$ and $\chi$ associated with hidden local and scale symmetries. The role of the topology change is to endow what could be identified as Kohn-Sham-type ``density-functional" structure in the parameters of the effective Lagrangian that are supposed to capture the topological structure of QCD variables in dense medium.

In this approach, there are neither explicit quark degrees of freedom nor  strangeness flavor as  in the standard approaches such as e.g., in \cite{baymetal,alford} and in other variations with bag models~\cite{MIT}.  Nonetheless,  up to today, the approach works with no serious tension with empirical data.  
It is  possible of course that  there be  corrections  to the approximations made --  given the admittedly drastic oversimplification  -- that could,  quantitatively though not qualitatively, modify the results. There is however one serious potential obstruction to G$n$EFT.  Should future measurements map out precisely the behavior of the sound velocity in the range of density $3\leq n/n_0\leq 7$ and falsify the precocious onset of,  and the convergence to,  the PC sound velocity, then that would bring a serious obstruction to the notion of the emergent symmetries, particularly hidden scale symmetry distinctive of the theory. That would then ``torpedo"  the G$n$EFT.   If however it is not ``torpedoed,"  then our approach with the encoded  ``duality" to QCD in approaching the chiral phase transition as well as confinement as argued recently by string-theory-inclined theorists~\cite{zohar,karasik1,karasik2,kitano,KKYY}  will bring a totally new perspective to nuclear physics, a paradigm almost totally foreign to nuclear theories. 

Most interesting future direction would be to map the ``generalized" sheet structure of Chern-Simons QFT in the topological sector conjectured above to an improved G$n$EFT phrased in Wilsonian-type Fermi-liquid theory more powerful and realistic than what has been achieved so far -- in terms of the half-skyrmion phase -- in the topological sector. It would offer a clear  resolution of the dichotomy problem and escape the possible obstruction to the G$n$EFT.

\subsection*{Acknowledgments}
We are grateful for collaborations and/or  discussions with Tom Kuo,  Hyun Kyu Lee, Maciej Nowak, Won-Gi Paeng,  Sang-Jin Sin and Ismail Zahed during and after the WCU/Hanyang Project.

\end{document}